\documentclass[manuscript]{aastex631}

\usepackage{amsmath,amssymb,bm}
\newcommand{\Bep}{B_\mathrm{ep}}
\newcommand{\Bet}{B_\mathrm{et}}
\definecolor{forestgreen}{rgb}{0.10, 0.50, 0.10}

\usepackage{cleveref}
\usepackage{color}           % For color text: \color command
\definecolor{DarkGreen}  {rgb}{0.10, 0.50, 0.10}
\definecolor{ForestGreen}{rgb}{0.0 , 0.61, 0.33} 
\definecolor{DarkMagenta}{rgb}{0.50, 0.0 , 0.50}
\definecolor{Bittersweet}{rgb}{0.75, 0.31, 0.09}
\definecolor{BlueViolet} {rgb}{0.28, 0.22, 0.57}
 \newcommand{\G}[1]{{\color{DarkGreen}{#1}}}

\received{\today}
\revised{ }
\accepted{ }
\shorttitle{Reconnection restraining flux rope eruption}
\shortauthors{J. Chen et al.}

\begin{document}

\title{A Model for Confined Solar Eruptions Including External Reconnection}

\author[0000-0003-3060-0480]{Jun Chen}
%\email{el2718chenjun@nju.edu.cn}
\affil{School of Astronomy and Space Science, Nanjing University, Nanjing 210093, China\\xincheng@nju.edu.cn}
\affil{Key Laboratory of Modern Astronomy and Astrophysics (Nanjing University), Ministry of Education, Nanjing 210093, China\\}

\author[0000-0003-2837-7136]{Xin Cheng}
\affil{School of Astronomy and Space Science, Nanjing University, Nanjing 210093, China\\xincheng@nju.edu.cn}
\affil{Key Laboratory of Modern Astronomy and Astrophysics (Nanjing University), Ministry of Education, Nanjing 210093, China\\}

\author[0000-0002-5740-8803]{Bernhard Kliem}
\affiliation{Institute of Physics and Astronomy, University of Potsdam, Potsdam 14476, Germany}

\author{MingDe Ding}
\affil{School of Astronomy and Space Science, Nanjing University, Nanjing 210093, China\\xincheng@nju.edu.cn}
\affil{Key Laboratory of Modern Astronomy and Astrophysics (Nanjing University), Ministry of Education, Nanjing 210093, China\\}

\begin{abstract}
The violent disruption of the coronal magnetic field is often 
observed to be restricted to the low corona, appearing as a confined 
eruption. 
The possible causes of the confinement remain elusive. Here, 
we model the eruption of a magnetic flux rope in a quadrupolar active region, 
with the parameters set such that magnetic X-lines exist
both below and above the rope. This facilitates the onset
of magnetic reconnection in either place but with partly opposing effects on the eruption.
The lower reconnection initially adds poloidal 
flux to the rope, increasing the upward hoop force and supporting
the rise of the rope. However, when the flux of the magnetic side lobes enters the lower reconnection, 
the flux rope is found to separate from the reconnection site and the flux accumulation ceases. 
At the same time, the upper reconnection begins to reduce the poloidal 
flux of the rope, decreasing its hoop force; eventually this cuts the rope completely. 
The relative weight of the two reconnection processes is varied in the model, 
and it is found that their combined effect and the tension force of the overlying field confine the eruption 
if the flux ratio of the outer to the inner polarities exceeds 
\G{a threshold, which is $\sim\!1.3$ for our Cartesian box and chosen parameters}. 
We hence propose that external reconnection between an erupting flux rope and overlying flux 
can play a vital role in confining eruptions.
\end{abstract}


\section{Introduction}\label{s:intro} 
Coronal mass ejections (CMEs) and flares are the two most violent energy release phenomena in the solar atmosphere. They are believed to be caused by the same process in essence, i.e., the eruption of a flux rope, which is defined as a set of twisted field lines around a central axis \citep[e.g.,][]{cheng2017,Patsourakos2020}. Nevertheless, the eruption of a flux rope does not always produce a CME. 
Based on the statistics of \citet{nindos2015}, 
about 45\% of \G{all} flares above M1-class are not accompanied by CMEs. However, even for 
flares without a CME, an erupting flux rope can \G{often} be observed, 
although it is eventually confined to the low corona. 
Moreover, such failed rope eruptions present an early kinematic evolution similar to successful ones 
\citep{cheng2020apj,huang2020apjl}.

The observation of a failed filament eruption in 
\citet{ji2003apj} spawned a strong interest in the possible causes of the 
confinement.
This particular event can be modeled as a kink-unstable flux rope in the stability domain of the torus instability \citep{Torok&Kliem2005, Hassanin2016apj}. However, since the helical kink instability appears to occur only in a minority of solar eruptions, their 
success or failure 
is often discussed in the framework of the properties of the torus instability \cite[TI,][]{Kliem&Torok2006}, whose threshold is given by 
a critical decay index $n_\mathrm{c}$. The decay index $n$ describes how fast the external poloidal field, $\Bep(R)$ (often simply referred to as the background field), declines with height, 
\begin{align}
n:=-\frac{\mathrm{d}\ln \Bep(R)}{\mathrm{d}\ln R},
\label{e:n_R}
\end{align}
where 
$R$ denotes 
the distance of the rope axis to the center of an assumed approximately toroidal rope. 
In the simplest case of a nearly toroidal flux rope shape and zero external toroidal (shear/guide) field, $\Bet=0$, the threshold is near its canonical value $n_\mathrm{c}=1.5$, but varying parameters, in particular the flux rope geometry and shear field strength, cause it to vary in the range $n_\mathrm{c}\sim1\mbox{--}2$ 
\citep{Kliem&Torok2006,Olmedo2010,Demoulin2010}. 
For $n>n_\mathrm{c}$ the rope is torus unstable and erupts. If this condition is fulfilled along the whole path of the rising rope, the eruption can be successful. 
This has been supported by a number of case and statistical studies \citep[e.g.,][]{guoyang2010apjl,cheng2011apjl,sunxudong2015,wangdong2017}. Nevertheless, it was found 
that torus-unstable rope eruptions may also suffer from failure if the decay index height profile, $n(h)$, possesses a sufficiently deep minimum, such that a torus-stable height range with $n<n_\mathrm{c}$ lies above a torus-unstable height range \citep{guoyang2010apjl}, or if the flux rope rotates strongly \citep{zhouzhenjun2019}. That is to say that the occurrence of torus instability is not a sufficient condition for a successful eruption.

Except for the decay property of the background field, the success or failure of an eruption is also influenced by other factors. Numerically and with laboratory experiments, it was revealed that a strong guide field component of the overlying field, $\Bet>\Bep$, is able to confine an erupting flux rope 
\citep{Torok&Kliem2005, Myers2015nature}. 
The cases of an upper torus-stable height range and a strong shear/guide field are often jointly referred to as configurations with a too strong overlying flux, and this is widely considered to be the most common reason for the confinement.
Moreover, the twist of the 
rope was found to be another decisive factor to influence the eruption \citep{Myers2015nature,liurui2016apj}. Based on careful analyses of a data-driven magnetohydrodynamic (MHD) simulation, \citet{zhongze2021nc} proposed that the non-axisymmetry of the rope is an additional critical factor to constrain its eruption.

Inspired by previous observations and simulations, in which the confinement of
eruptions could result from a too strong overlying field \citep{DeVore2008apj} or from external reconnection between the erupting flux
and the overlying field \citep{Netzel2012, Hassanin2016apj, Kumar2023apj},
we here investigate 
the joint action of these related effects 
in the specific topology of a quadrupolar source region, which facilitates external reconnection.
The rest of the paper is arranged as follows: 
in Section~\ref{s:method}, the numerical model is detailed. 
In Section~\ref{s:results}, we present the results of the simulations, 
followed by a summary and discussions in Section~\ref{s:summary}.

\section{Method}\label{s:method} 

\subsection{Initial magnetic field}\label{ss:field} 
\begin{figure}[ht]
    \centering
    \includegraphics[width=0.8\linewidth]{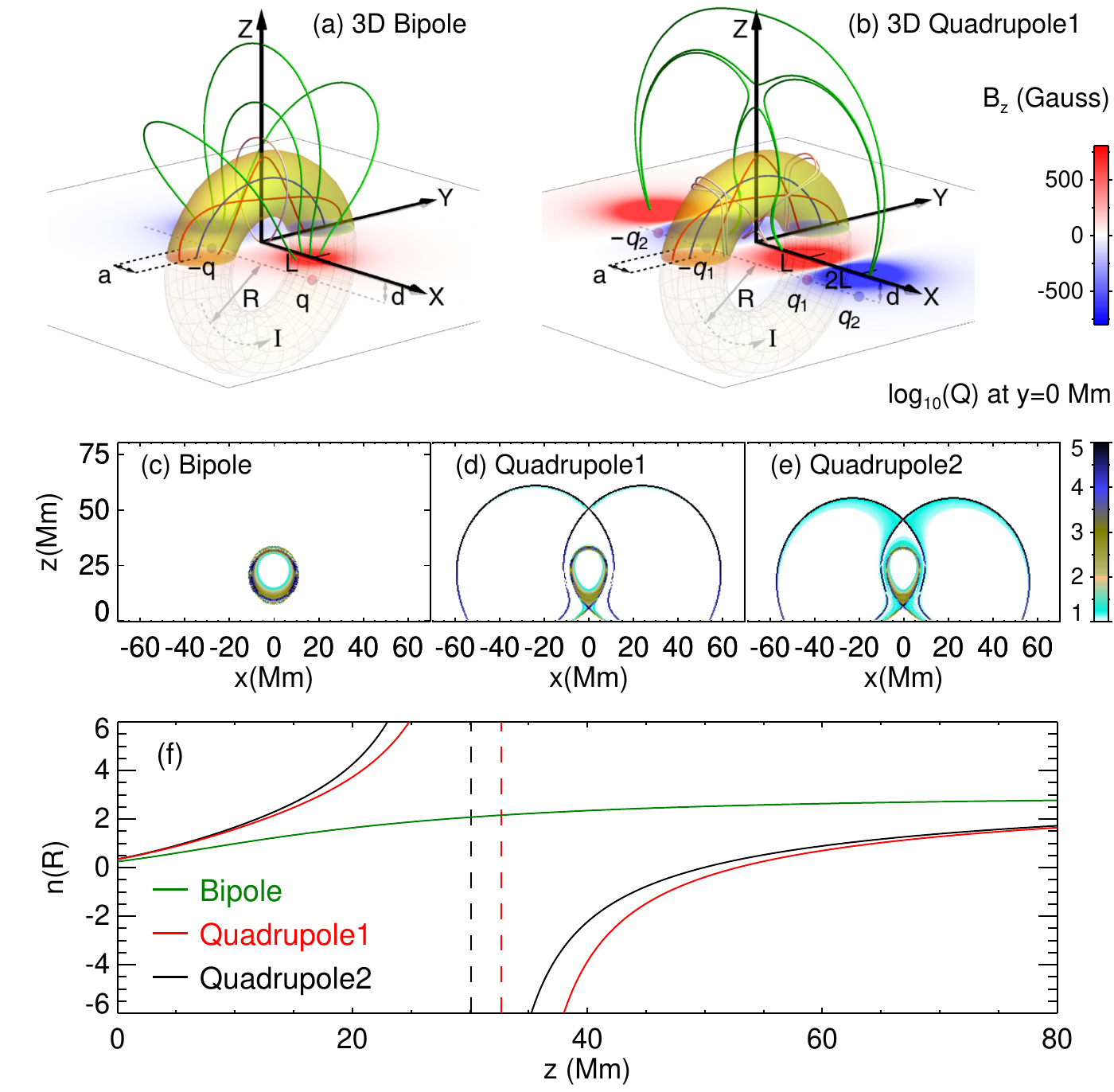}
    \caption{Initial configurations of the simulations. (a) Bipole case, $q_2=0$. (b) Quadrupole1 case with $q_2=-4 \times 10^{13}~\text{T}\,\text{m}^2$. 
    The blue field line is the magnetic axis, the transparent yellow tube shows the toroidal current channel. 
    (c)--(e)
    % Traces of high 
    Distributions of 
    squashing factor $Q$ in the plane $y = 0$, characterizing the magnetic topology, including Quadrupole2.
    \G{
    (f) $n(R)$ of the initial configurations, plotted vs.\ height $z=R-d$; dashed lines mark the 
        heights where $n(R) \to \pm\infty$. 
    }
}
\label{fig:initial_configuration}
\end{figure}
The classic model of a force-free flux rope by \cite{TD99} (hereafter TD99)
is illustrated in Figure~\ref{fig:initial_configuration}(a).
A toroidal ring current of  major radius $R$ and minor radius $a$ is centered at $(0,\, 0,\, -d)$. 
A pair of magnetic charges $\pm\, q$ at $ (\pm\, L,\, 0,\, -d)$ provides the external poloidal field.
For the balance between the upward hoop force and the downward strapping force from the external field, 
the equilibrium current $I$ is given by 
\begin{align}
I=\frac{8\, \pi\,q\,L\,R\,(R^2+L^2)^{-3/2}}{\mu_0\,[\ln(8\,R/a)-3/2+l_i/2]},
\label{eq:Iq}
\end{align}
where $l_i$ is the internal self-inductance per unit length of the
tube \citep{Shafranov1966}. This quantity depends weakly on the current distribution in the ring. 
For simplicity, we set $\Bet=0$.

In this work, we modify the TD99 model to set a flux rope in a quadrupolar active region (Figure \ref{fig:initial_configuration}(b)).
This is constructed by adding a second pair of magnetic charges with the strength of 
$\pm\,q_2$ at $(\pm\G{L_2},\, 0,\, -d)$ \G{with $L_2=2L$.} 
In order to yield the same strapping field strength at the geometrical torus axis 
\G{(at distance $R$ from torus center)} as \G{for the bipole,} 
% as in Figure~\ref{fig:initial_configuration}(a), 
the strength of the inner pair of charges is adjusted to 
\begin{align}
q_1= q-q_2\G{\frac{L_2}{L}} \left(\frac{R^2+L^2}{R^2+\G{L_2}^2}\right)^{3/2}.
% q_1= q-2\, q_2\, \left(\frac{R^2+L^2}{R^2+4\,L^2}\right)^{3/2}.
\label{eq:q1}
\end{align}
The flux from the inner pair yields a downward force, and the flux from the \G{outer} pair yields an upward force.

We set $R=27.5$~Mm, $a=11.1$~Mm, $d=7.5$~Mm, $L=25$~Mm, $q=10^{13}~\text{T}\,\text{m}^2$, and $l_i=0.5$.
Three initial configurations are created by setting $q_2$ to $\{0, -4, -5\} \times 10^{13}~\text{T}\,\text{m}^2$, and the corresponding $q_1$ are derived from Equation \eqref{eq:q1}, so that $- q_2/q_1=$ 0, 1.246, 1.329, respectively. 
These runs are denoted with Bipole, Quadrupole1, and Quadrupole2, respectively. 
As shown in Figure~\ref{fig:initial_configuration}, the quadrupolar configurations possess two X- (null) lines, one above and one below the flux rope, which facilitate the onset of magnetic reconnection. The corresponding heights of the apex of the upper X-line are 50.3 and 45.2~Mm.

\subsection{Numerical model}\label{ss:numerics} 
Before the simulation, a normalization is performed referring to the values at the apex of the geometric toroidal axis $(0,\,0,\,R-d)$.
We take the height $R-d$, initial field strength $B_0$, density $\rho_0$, corresponding Alfv\'en speed $V_A=B_0/ \sqrt{ \mu_0\,\rho_0}$ and corresponding Alfv\'en time $\tau_A=(R-d)/V_A$ at this site as the units of the corresponding variables. For example, for $V_A=1000$~km\,s$^{-1}$, we have $\tau_A=20$~s.
The computations are performed in a Cartesian cubic box of
\G{$[-640,640]\times[-640,640]\times[0,1280]$}~Mm.

We integrate the normalized ideal MHD equations neglecting gravity and thermal pressure:
\begin{align}
\partial_t\, \rho&= -\bm{\nabla\cdot}(\rho\,\bm{u})\,, \label{eq_rho}\\
\partial_{t}\,(\rho\,\bm{u})&=-\bm{\nabla \cdot} (\rho \, \bm{u}\, \bm{u})
+\bm{\nabla \cdot \mathsf{T}}  +\bm{J\times B}\,,  \label{eq_mot}\\
\partial_{t}\,\bm{B}&= -\bm{\nabla\cdot}(\bm{u\,  B-B\,  u})\,, \label{eq_ind}
\end{align}
where $\bm{J}\equiv\bm{\nabla\times B}\, $is the current density, $\bm{\mathsf{T}}\equiv{R_e}^{-1}\,\rho\,
[\bm{\nabla\,  u}+(\bm{{\nabla\,  u}})^T-(2/3\,\bm{\nabla\cdot u })\, \bm{\mathsf{I}}]$ is the viscous stress tensor,  $\bm{\mathsf{I}}$ is the second order unit tensor, $^T$ denotes the transposition for a second-order tensor, and $R_e$ denotes the fluid Reynolds number. 
Closed boundaries are applied ($\bm{u} = \bm{0}$, at all boundaries), 
resulting in an invariant normal magnetogram component ($ \partial\, B_z / \partial\, t|_{z=0} = 0$).

\Cref{eq_rho,eq_mot,eq_ind} are integrated by the modified Lax-Wendroff scheme described in \cite{Torok&Kliem2003}. In place of the diffusive Lax step, 
artificial smoothing \citep{Sato&Hayashi1979} is applied to $\rho$ through the substitution 
$\rho_i \to (1-c_\rho)\, \rho_i + c_\rho/6\, \sum_j \rho_j$, where $j$ are the 6 neighbor grid points of $i$. 
This is similar in structure to the Lax term, which has $c_\rho=1$, but far less diffusive for small values of $c_\rho$. 
 This smoothing is also applied to $\bm{u}$ and $\bm{B}$. The latter introduces numerical resistivity, which facilitates magnetic reconnection.
We set $c_\rho= c_u= 0.01-0.1$ (exponentially decreasing with height in [0, 100]~Mm, and staying at 0.01 in the region above), and choose a small, 
uniform \G{$c_B = 0.001$} to ensure that magnetic diffusion is not significant outside of the reconnection regions.
% \G{
% The numerical errors of $\nabla\cdot\bm{B}$ were eliminated through the diffusive treatment as proposed by 
% \cite{Dedner2002}.}
\G{The nonzero $\nabla\cdot\bm{B}$ resulting 
from the finite differences is kept small by the standard diffusive treatment following \citet{Dedner2002}.} 
The initial density is set to $\rho(\bm{x},\, t=0)=|\bm{B}(\bm{x},\, t=0)|^{3/2}$
(see, e.g., \cite{Torok&Kliem2005} for a discussion of this choice). 
The initial velocity  is set to $\bm{u}(\bm{x},\, t=0)=\bm{0}$. 

3D magnetic reconnection preferentially takes place 
where a large gradient of magnetic connectivity is present.
Such connectivity change can be quantified by the squashing factor $Q$ \citep{Titov2002,Titov2007}. 
Separatrices are located where $Q = \infty$, 
quasi-separatrix layers (QSLs) are located where $Q \gg 2$.
The distribution of $Q$ in 
\G{the midplane of the configuration, $\{y=0\}$,} 
% selected planes
is here computed following \cite{FastQSL}. 
Two separatrices (QSLs) intersect with each other in a separator (quasi-separator or hyperbolic flux tube \citep[HFT;][]{Titov2002}). Such intersections, jointly referred to 
as ``(quasi-) separators'' of the magnetic field in the following, are the favorable sites for 3D magnetic reconnection \citep{Priest2000,Pontin2011}. 
These topological structures allow us to
quantify the 3D reconnection processes at the different locations and their temporal evolution. Because we have set $\Bet=0$, our quadrupolar configurations initially contain true separators, the X-lines, which would change to HFTs if $\Bet\ne0$.

\section{Results}\label{s:results} 

\begin{figure}[ht]
    \centering
    \includegraphics[width=0.7\linewidth]{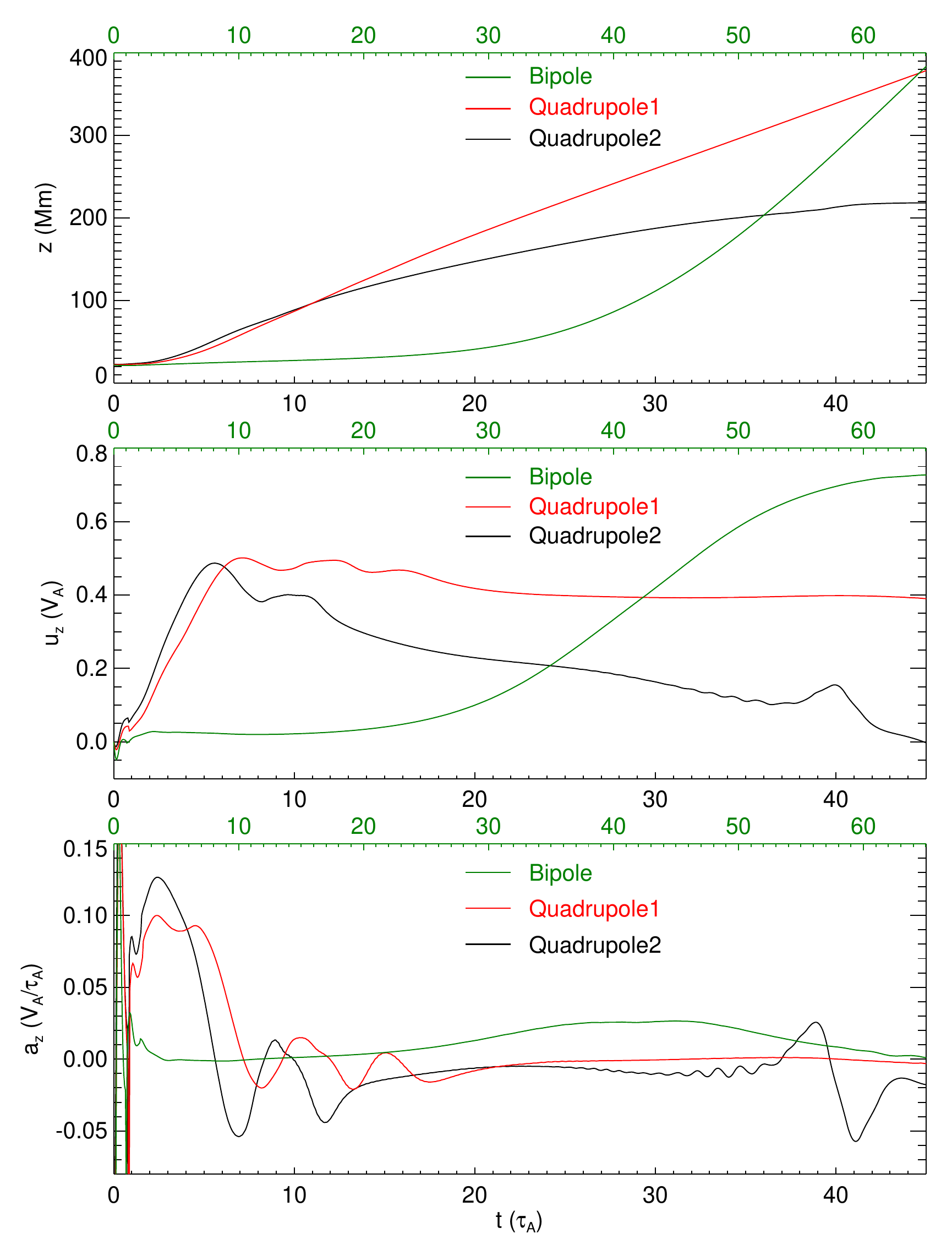}
    \caption{
    Height-time (upper), velocity-time (middle) and acceleration-time (lower) profiles of the apex point of the flux rope axis in the three runs.
    The run 
    Bipole (green profiles) is scaled with the green $t$-axis, other runs are scaled with the black $t$-axis. }
\label{fig:cfls}
\end{figure}

The rise profiles of the magnetic axis' apex point are shown in Figure \ref{fig:cfls} for the three runs.
The decay index of the external poloidal field at the initial magnetic axis of Bipole and Quadrupole1--2 configurations are  1.73, 4.65, and 5.77, respectively.
For the Bipole case, the initial flux rope is \G{only} 
slightly above the marginal\G{ly} unstable state, therefore it takes a 
relatively long time to erupt. 
\G{The Quadrupole1--2 cases are not only intially positioned much further into 
the unstable domain of parameter space, but their} 
external poloidal field 
\G{continues to} decrease \G{much faster} 
with height  
up to the null line \G{and field reversal, where $n(R)$ has a pole} (Figure~\ref{fig:initial_configuration}(f)). 
\G{Consequently, their instability commences immediately and develops stronger.} 
\G{It is worth noting that the relative magnetic helicity is largest for the Bipole and far smaller for Quadrupole1--2, with the flux-normalized values being $H_\mathrm{m}/F^2=-0.15$, $-0.004$, and $-0.002$, respectively. The small values of the quadrupolar cases result from the opposite relative helicities between the flux rope and the oppositely directed inner and outer bipole fields. 
% , which largely cancel each other.  
}

\begin{figure}[ht]
    \centering
    \includegraphics[width=0.8\linewidth]{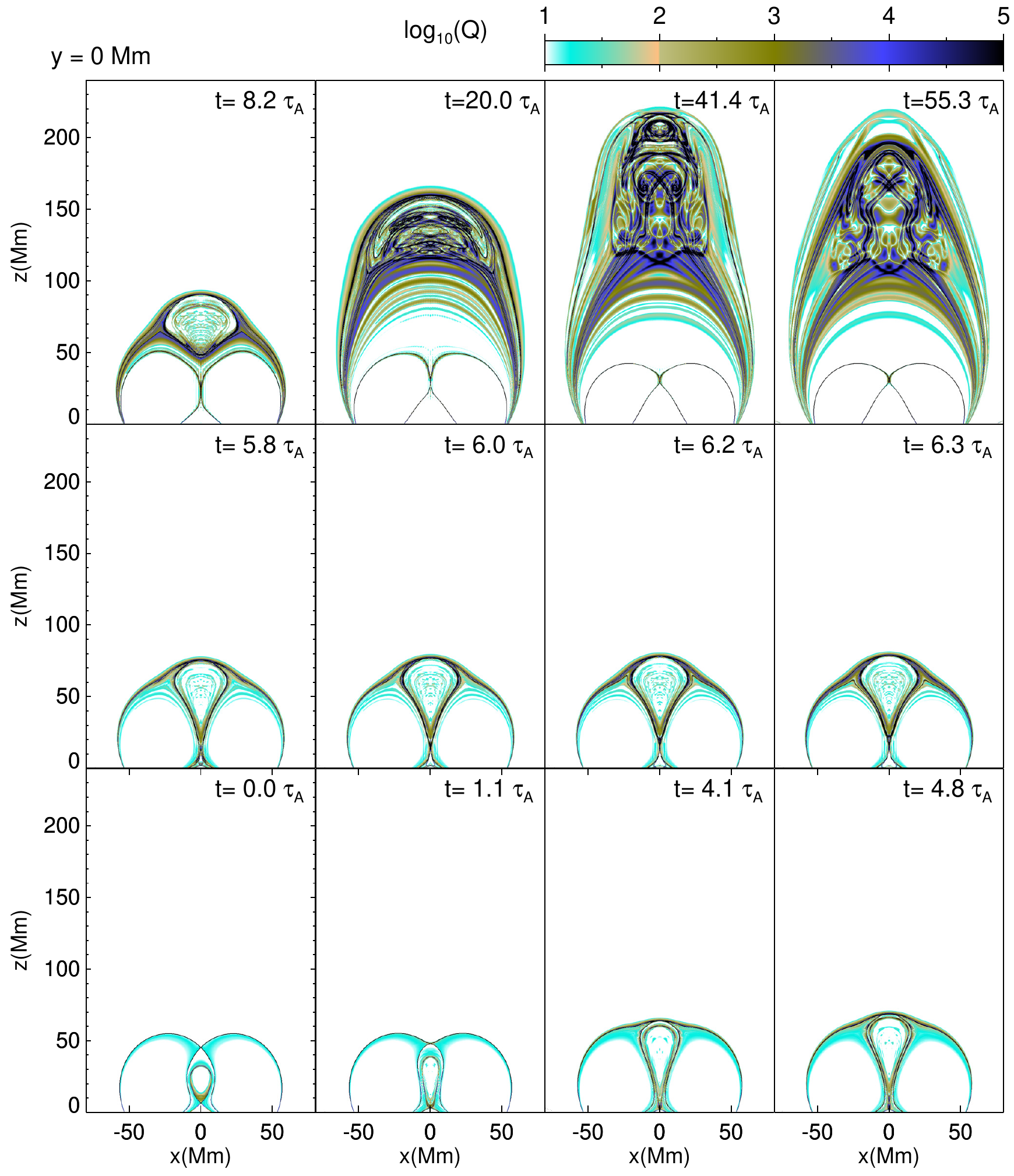}
    \caption{
    Squashing factor $Q(x,0,z)$
    at different times for the run Quadrupole2, showing the evolution of the magnetic topology. 
    \G{An animation is available online to show the evolution of the Squashing factor $Q(x,0,z)$ distribution during $t \in $ [0, 55.3]~$\tau_A$. The duration of the animation is 26 s.}
    }
\label{fig:qsl3}
\end{figure}

\begin{figure}[ht]
    \centering
    \includegraphics[width=0.7\linewidth]{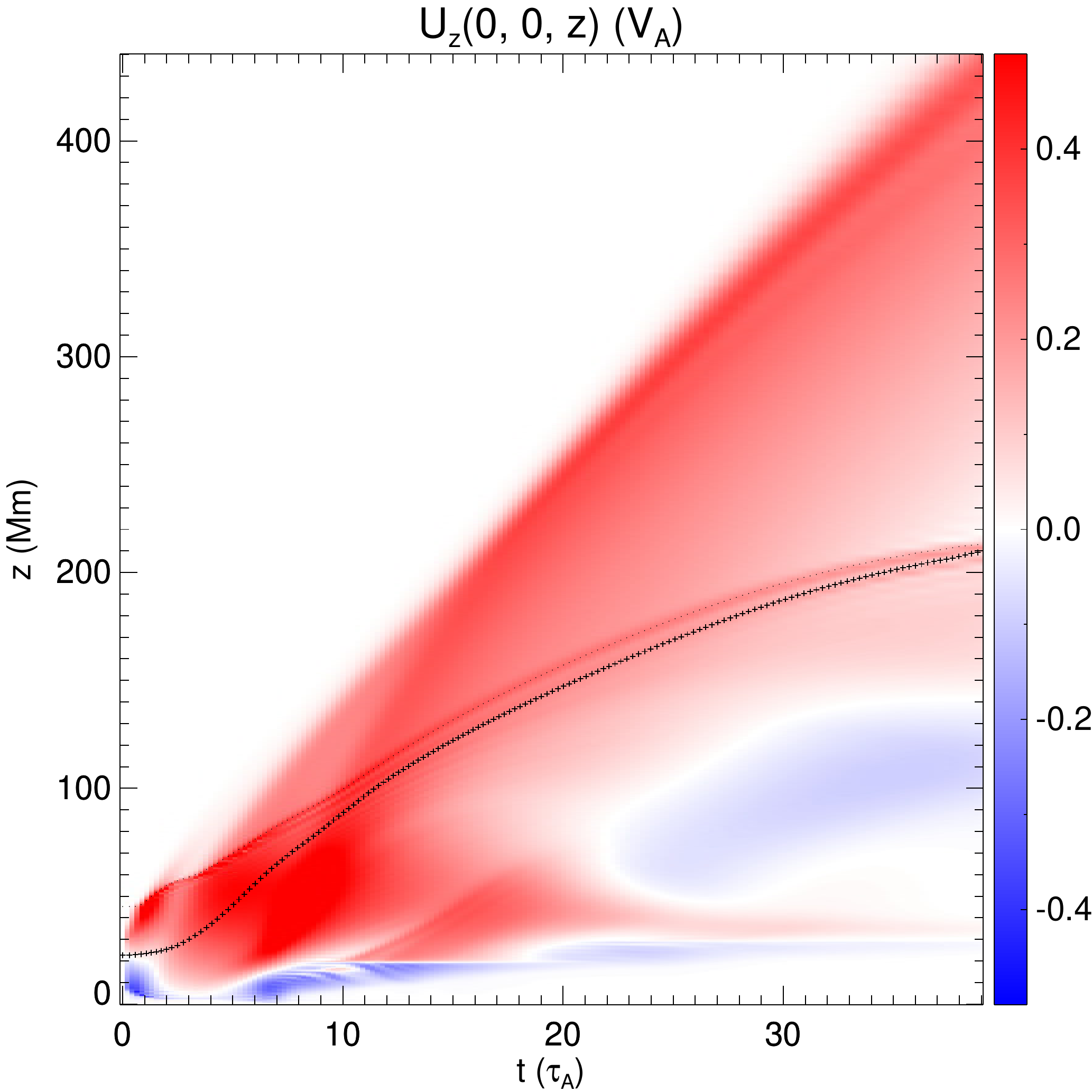}
    \caption{Time-distance plot of $U_z(0, 0, z)$ for the run Quadrupole2. Plus signs mark the position of the magnetic rope axis.
    \G{A} Dotted line, where $B_x = 0$, trace\G{s} the \G{motion} of the \G{apex} of the 
    upper (quasi-) separator.}
\label{fig:uzt}
\end{figure}

\begin{figure}[ht]
    \centering
    \includegraphics[width=0.84\linewidth]{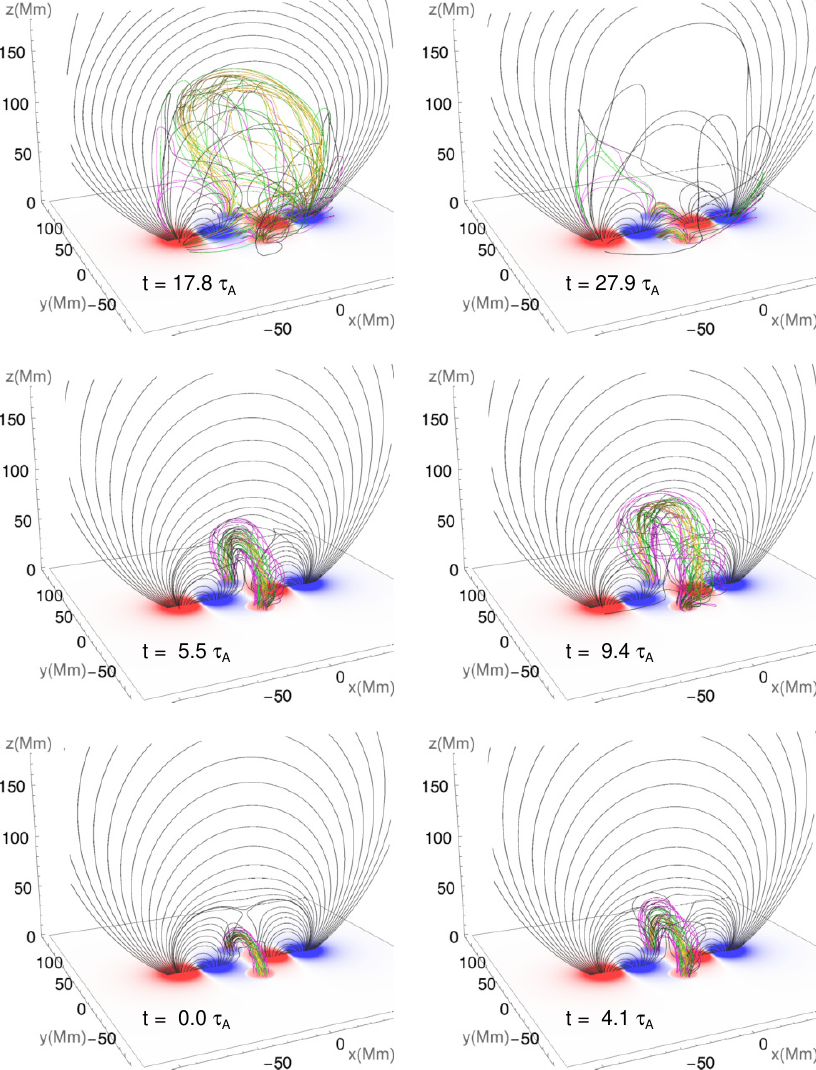}
    \caption{
    Temporal evolution of the field lines for the run Quadrupole2.
    The color scale of $B_z$ is the same as
    Figure \ref{fig:initial_configuration}.
    \G{An animation is available online to show the evolution of the magnetic field lines during $t \in $ [0, 55.3]~$\tau_A$. The duration of the animation is 14 s.}
    }
\label{fig:flb}
\end{figure}

We first focus on the evolution of Quadrupole2. 
Two mechanisms drive the acceleration: 
torus instability and ``flare'' reconnection in the vertical (``flare'') current sheet that forms from the initial X-line under
the flux rope. This reconnection feeds poloidal magnetic flux---the strapping flux in the center lobe---into, and generates an upward outflow toward, the flux rope. 
However, the flux feeding only takes a short period, up to \G{$t\approx6~\tau_A$}. 
% During the following $\sim3~\tau_A$,
\G{Subsequently,}
the flux rope axis \G{quickly} separates from
the lower reconnection region (\G{within $\sim3\mbox{--}4~\tau_A$; see} Figure~\ref{fig:qsl3}). 
The separation is also obvious from the increasing distance between 
the magnetic rope axis and the upper edge of the upward reconnection outflow  \G{in} Figure~\ref{fig:uzt}. The reason for the separation and the subsequent decline of the flare reconnection (Figure~\ref{fig:uzt}) is the change from strapping-flux to side-lobe-flux reconnection, which happens when the bounding separatrices
meet at the flare current sheet (see Figure~\ref{fig:qsl3} at 
\G{$t=6.0~\tau_A$}). 
The reconnected side-lobe flux does not wrap around the erupting flux rope, but rather forms high-lying loops below the rope and above the side lobes (Figure \ref{fig:flb}, 
\G{$t=9.4~\tau_A,\,17.8~\tau_A$}). This reconnected flux separates the flux rope from the region of flare reconnection, terminating the support of the eruption by the flare reconnection through both flux accretion and momentum transfer by the reconnection outflow jet.

Figure~\ref{fig:uzt} shows a high gradient of $U_z$ at the upper (quasi-) separator, 
corresponding to an inflow in the reference frame tied to 
the (quasi-) separator which rises with the flux rope. 
This indicates that 
reconnection also acts above the rope in a horizontal current sheet forming from the upper initial X-line
from the very beginning of the simulation. The upper reconnection reduces the constraining overlying flux, 
and initially transfers it to the side lobes, joint with the reconnected strapping flux from the center lobe,
as in the breakout model. However, when all strapping flux is reconnected,
the upper reconnection begins to involve the flux rope, building up a connection of each rope footprint to the outer ambient flux sources.
At the same time, the side lobes meet in the flare current sheet (Figure~\ref{fig:qsl3}). 
The resulting erosion of the rope flux corresponds to a
reduction of the toroidal current $I$, which is proportional to the poloidal flux of the rope. 
The strapping force in the rope is proportioned to $I$, while the upward hoop force is proportioned to $I^2$. 
Thus, the upper reconnection with the flux rope weakens the net upward force in the rope
that drives the eruption.

Figure~\ref{fig:cfls} shows that the main
upward acceleration of the rope ends at \G{$t\approx5.2~\tau_A$}, 
shortly before all strapping flux is reconnected and the upper reconnection and the separation of the rope begin. 
The
tension force of the overlying flux rooted in the outer polarities
decelerates the rope in this interval, but the major deceleration, 
leading to the confinement, 
happens \G{in a much longer} subsequent \G{period} when all three effects act jointly (Figure~\ref{fig:cfls}, middle).

A transitory amplification of the reconnection outflows occurs when the side lobes join the lower reconnection because of their higher flux density ($|q_2|>q_1$). 
This enhances the upward tension force of the upper outflow, strongly acting on
the flux rope around $t=10~\tau_A$ (Figure~\ref{fig:uzt}). The resulting
transitory second acceleration of the rope 
remains minor in the overall evolution of the eruption (Figure~\ref{fig:cfls}). 

In the early phase, 
both lower and upper reconnection are proceeding simultaneously
 (Figure \ref{fig:flb}, \G{$t = [4.1~\tau_A,\,5.5~\tau_A$]}).
While the lower reconnection decouples from the flux rope after 
\G{$t\approx6~\tau_A$}, 
the upper reconnection acts strongly on the rope 
\G{during most of the deceleration phase} 
% throughout the simulation 
(Figure \ref{fig:flb}). 
This results in all rope flux being peeled off, eventually destroying the rope 
(Figure \ref{fig:flb}, \G{$t = 27.9~\tau_A$}).

The run of Quadrupole1 also 
shows the separation of the flux rope from the lower 
reconnection region, only slightly later, also leading to deceleration. 
This eruption is intermediate between the Bipole and Quadrupole2 runs. 
\G{It shows
%  a rise into
 a rise \G{above 19 times} of the initial height
% into the upper half of the box 
at a considerable speed 
($U_z\approx0.4\,V_\mathrm{A}$) with only a weak deceleration} (Figure~\ref{fig:cfls}). 
\G{With the given 
% (large) 
box, this eruption is ejective. Because the eruption shows a propagation phase dominated by inertia already from $\approx1/10$ of the box height, an even larger box is expected to yield the same result. 
On the other hand, an exponential phase, a typical characteristic of instability, is not clearly seen here (similar to Run Quadrupole2). Comparing the two quadrupolar configurations, it is obvious that} 
a stronger $q_2$ 
forms a lower null and provides a higher amount of \G{overlying} flux that can be reconnected at the upper null; then the effect of restraining the eruption 
\G{by the upper reconnection} is more significant. Additionally, the downward tension force of the overlying flux is higher. 
\G{We have further constrained the point of transition between confined and ejective behavior of the eruption in the quadrupolar configuration and found it to lie in the range $q_2\in 
[425,450]\times10^{13}$~T\,m$^2$, 
corresponding to $-q_2/q_1\in[1.27,1.29]$. 
This range depends on the parameters $d/R$, $a/R$, and $L/R$, as well as (weakly) on the numerical settings.} 

% \clearpage 
\section{Summary and Discussions}\label{s:summary} 
In this Letter, we present a model for confined 
solar eruptions 
in quadrupolar field configurations with a flux rope pre-existing in the central flux lobe. 
A new key feature of the model is the change in character of the upper reconnection 
in the current sheet forming from the (quasi-) separator 
above the erupting flux rope when all strapping flux in the center lobe has reconnected and the rope itself enters the reconnection. The upper 
reconnection supports the eruption initially, in full agreement with the breakout model, 
according to which the reconnection 
moves overlying flux to the side lobes, decreasing its restraining force 
\citep{Antiochos1999apj, DeVore2008apj, Karpen2012}. 
However, after all strapping flux in the center lobe has been removed, the upper reconnection erodes the flux rope, decreasing its upward hoop force
that drives the eruption and \G{it can} eventually \G{destroy} the rope.

At the same time, the flux of the side lobes enters the vertical current sheet under the rope.
The lower reconnection\G{, often referred to as ``flare reconnection'',} also changes its character at this point. 
The flux in the upward reconnection outflow then does no longer wrap about the erupting flux rope but rather forms simple loops above the side lobes. This results in a separation of the erupting flux rope from the lower reconnection region, so that the strengthening of the upward force by flux accretion and momentum transfer to the rope terminates, 
very soon after the flux rope has risen above the side lobes.

Thus, in eruptions from the central lobe of a quadrupolar configuration, 
both the upper and lower reconnection act against the eruption when all strapping flux in the center lobe has been removed. Jointly with the standard tension force of the overlying flux rooted in the outer polarities, 
this \G{can confine} the eruption.

The mechanism does not require the flux rope to exist before the onset of the eruption. 
Rather, it can work in the same way when the formation of the flux rope commences 
simultaneously with the eruption \cite[as, e.g., in][]{Mikic&Linker1994, Karpen2012, JiangC&al2021}.
 It does require that the strapping flux in the center lobe, i.e., 
the total flux in the center lobe minus the rope flux at the onset of the eruption 
(e.g, by onset of the torus instability), be smaller
than the overlying flux rooted in the outer polarities. 
This is guaranteed if $|q_2|>q_1$ but fulfilled also in a range of $|q_2|$ somewhat smaller 
than $q_1$ in the case of a pre-existing flux rope. However, 
to explain the confinement of eruptions at the typically observed heights in the \G{low to middle} corona, 
up to about $z\sim(3/4)\, R_\odot$ and $\sim$\,20 times the initial height \cite[e.g.,][]{Koleva&al2012}, 
\G{the ratio $|q_2/q_1|$ must not be too small. 
The runs} 
Quadrupole1 \G{and 2 and intermediate test runs} suggest 
\G{a threshold of $|q_2/q_1|\sim1.3$. The threshold depends on the parameters, 
primarily on $d/R$, which generally influences the stability properties of the TD99 flux rope, 
and on $L_2/L$, which determines the height profile of the flux overlying the flux rope and, hence, 
the amount of overlying flux jointly with $|q_2/q_1|$. For larger $L_2/L$, 
the field strength above the rope 
decreases less with increasing height, so that the threshold of $|q_2/q_1|$ is expected to decrease slightly. On the other hand, in spherical coordinates the field strength decreases faster, implying a somewhat higher threshold on the Sun.}

\G{The threshold value appears} consistent with the source-region properties of confined vs. eruptive flares in \citet{WangY&ZhangJ2007} and \citet{cheng2011apjl}. All 
confined events occurred in the central part of complex source regions suggestive of outer overlying flux at least as strong as the central flux around the erupting part of the polarity inversion line. All 
ejective events occurred near the periphery of the source region where no such strong outer overlying flux was present.

Our results are also consistent with the confinement \citep{DeVore2008apj} and 
success \cite[e.g.,][]{Lynch&al2008} of breakout eruptions that have used $|q_2|$ significantly larger (smaller) than $q_1$ and showed (did not show) reconnection of side-lobe flux under the erupting flux, respectively.
% [The confined breakout eruptions in \citet{DeVore2008apj} were found with $|q_2|$ significantly exceeding $q_1$ and showed reconnection of side-lobe flux under the erupting flux, while the ejective one in \citet{Lynch&al2008} used the opposite setting and no such reconnection was displayed. Similarly, 
The reconnection of side-lobe flux in the ejective breakout eruption in \citet{Karpen2012} 
begins just when the kinetic energy in the box stops rising (see their Figures~4 and 14); 
at this time the flux rope is already too high ($z\sim4\,R_\odot$) to be stopped by the remaining overlying flux
\G{, i.e., their flux ratio corresponding to our $|q_2/q_1|$ must be below the treshold for confinement}.

\G{As revealed in the Quadrupole1 run, the external reconnection might also responsible for the deceleration of CMEs in the high corona, which is usually ascribed to aerodynamic drag \citep[e.g.,][]{Vrsnak2004aa,shitong2015apj}. Another implication is the transfer of magnetic helicity and twist from the reconnecting flux rope to the ambient field. It provides a possible interpretation for the formation of large-scale flux ropes as indicated by large-scale filaments in the vicinity of sunspots \citep[e.g.,][]{guo2019apjl}.}

\section{Acknowledgements}
J.C., X.C., and M.D.D. are funded by National Key R\&D Program of China under grant 2021YFA1600504 and by NSFC grant 12127901. B.K. acknowledges support from the DFG and from NASA through grants 80NSSC19K0860, 80NSSC19K0082, and 80NSSC20K1274.

\end{document}